%% file: quantum.tex
\documentclass[twocolumn,pre,superscriptaddress,showpacs]{revtex4}
\usepackage{graphicx}
\usepackage{times}
\usepackage{amsmath}
\usepackage{amssymb}

\newcommand{\bra}[1]{\left\langle{#1}\right|}
\newcommand{\ket}[1]{\left|{#1}\right\rangle}

\begin{document}

\title{Optimized Broad-Histogram Ensembles for the Simulation of Quantum Systems}

\author{Stefan Wessel}
\affiliation{Institut f\"ur Theoretische Physik III, Universit\"at Stuttgart, 70550 Stuttgart, Germany}
\author{Norbert Stoop}
\affiliation{Theoretische Physik, Eidgen\"ossische Technische Hochschule Z\"urich, 8093 Z\"urich, Switzerland}
\author{Emanuel Gull}
\affiliation{Theoretische Physik, Eidgen\"ossische Technische Hochschule Z\"urich, 8093 Z\"urich, Switzerland}
\author{Simon Trebst}
\affiliation{Microsoft Research, Station Q, University of California, Santa Barbara, CA 93106, USA}
\author{Matthias Troyer}
\affiliation{Theoretische Physik, Eidgen\"ossische Technische Hochschule Z\"urich, 8093 Z\"urich, Switzerland}

\date{\today}

\begin{abstract}
The efficiency of statistical sampling in broad-histogram Monte Carlo simulations can be 
considerably improved by optimizing the simulated extended ensemble for fastest equilibration. 
Here we describe how a recently developed feedback algorithm can be generalized to find
optimized sampling distributions for the simulation of quantum systems in the context of the stochastic 
series expansion (SSE) when defining an extended ensemble in the expansion order.
If the chosen update method is efficient, such as non-local updates for systems undergoing
a second-order phase transition, the optimized ensemble is characterized by a flat histogram in the
expansion order if a variable-length formulation of the SSE is used.
Whenever the update method suffers from slowdown, such as at a first-order phase transition, the feedback algorithm shifts weight towards the 
expansion orders in the transition region, resulting in a
non-uniform histogram. 
\end{abstract}


\maketitle

\section{Introduction}
\label{Sec:Intro}

Quantum Monte Carlo techniques are widely employed to study equilibrium properties of quantum systems that do not suffer from the infamous negative-sign problem which 
renders any efficient simulation impossible \cite{SignProblem}. 
Over the last decade many technical advances including the adaptation of 
non-local update techniques 
\cite{LoopUpdates,Henelius,Wiese,Worms,OperatorLoops,DirectedLoops,GenDirLoop}
from their classical counterparts \cite{ClusterUpdates} and the implementation
of continuous-time algorithms \cite{Wiese,Worms,ContinuousTimeLocalUpdates} 
-- which only later have been adapted to classical systems \cite{ClassicalWorms} -- 
have made quantum Monte Carlo (QMC) simulations almost as powerful as classical 
simulations, despite their usually more complex formulation and implementation.

Non-local update techniques efficiently overcome the critical slowing-down at
many second order-phase transitions \cite{NonLocalUpdates,NaokiReview}, but do not help to overcome
the problem of tunneling out of metastable states at first-order phase transitions or
the sampling of rough energy landscapes. 
To alleviate these sampling problems for systems with competing, but well-separated 
states a number of techniques have been developed that all aim at broadening 
the range of sampled reaction coordinates such as parallel tempering / replica 
exchange methods \cite{ParallelTempering} or 
alternative extended ensemble techniques
which include multicanonical sampling \cite{MuCa}, broad-histogram sampling
\cite{BroadHistograms} and the Wang-Landau algorithm \cite{WangLandau}.
Similar approaches have
been formulated also for QMC
simulations~\cite{SandvikParallelTempering,SandvikQuantumParallelTempering,QWL,NaokiTM}.
While most of the extended ensemble methods sample a flat histogram for a 
given reaction coordinate, it has been pointed out that flat-histogram sampling
generally leads to suboptimal scaling behavior \cite{FlatHistograms}. If applied
to second-order phase transitions this slowing-down can be overcome by 
introducing non-local update techniques \cite{ClusterExtended}. In the more
general case, it has recently been demonstrated that even for local update 
techniques this slowing-down can be overcome by optimizing the sampled statistical
ensemble \cite{OptimizedEnsembles}. The idea is to identify bottlenecks of the
simulation by measuring the local diffusivity of the random walk in a reaction 
coordinate and to systematically feedback this information to shift resources
(and statistical weight) towards those regions where the local diffusivity is 
suppressed. The resulting histograms are {\em non-uniform} sampling distributions 
which are typically tailored to the underlying problem.
The feedback algorithm has been applied to a range of classical systems with 
rough energy landscapes including frustrated magnets \cite{SpiralSpinLiquids}, 
small proteins \cite{Proteins} and dense liquids \cite{DenseLiquids}, and
has recently been extended to optimize the simulated temperature set in 
parallel tempering / replica exchange simulation schemes 
\cite{OptimizedTempering,NadlerHansmann}.

In this manuscript, we discuss how the ensemble optimization technique can be
applied to efficiently sample quantum systems in the context of a stochastic
series expansion (SSE) formulation. Our approach extends previous ideas of adapting the
Wang-Landau algorithm to quantum systems  by observing that
the stochastic series expansion performs a one-dimensional random walk in
expansion orders, which similar to the classical algorithm can be 
biased using a generalized density of states (in expansion orders)~\cite{QWL}. 
It has been demonstrated that 
sampling an extended 
ensemble can significantly improve the sampling at  first-order
quantum phase transitions~\cite{SandvikQuantumParallelTempering,QWL}.
Obtaining a QMC estimate for the generalized density of states  allows 
for the direct calculation of the 
free energy, internal energy, entropy and specific heat for a wide range of
temperatures.
Here we apply the feedback algorithm to study and further improve the sampling
efficiency of such extended ensemble quantum simulations. 

For thermal second-order phase transitions where typically non-local update algorithms 
are efficient we find that using non-local updates and sampling a flat histogram in expansion 
orders is already optimal.
However, we find a dependency on the underlying representation of the operator string in 
the stochastic series expansion which we discuss in detail. We show that the commonly 
employed fixed length vector representation gives inferior results to a variable length list 
representation. It is only for the latter the optimal ensemble is just a flat histogram in the 
expansion orders.

For first-order phase transitions, both thermal and quantum, we find a significant
reweighing when applying the feedback technique resulting in model-specific
histograms. We show that these non-uniform sampling distributions further 
improve the simulations when compared to the flat-histogram approach, and 
result in an unbiased distribution of statistical errors.

This paper is organized as follows: In section \ref{Sec:SSE} we outline 
the stochastic series expansion algorithm and discuss both fixed length
and variable length representations for the sampled operator string.
We introduce modifications to this algorithm to sample extended statistical ensembles
in section \ref{Sec:ExEn} and explain the adaptive feedback algorithm that finds the
optimized broad-histogram ensemble. Finally, we discuss application of these
sampling techniques to a number of quantum spin systems in section \ref{Sec:Examples}
and give systematic comparisons to flat-histogram sampling methods.
We close with a summary of our results and in an appendix give a detailed analysis
of the variable length representation for the operator string and the respective
update algorithms.

\section{Stochastic series expansions}
\label{Sec:SSE}

The fundamental problem for Monte Carlo simulations of quantum systems is that the 
partition function is not a simple sum over classical configurations but an
operator expression
\begin{equation}
  Z = {\rm Tr} \exp(- \beta H) \,, 
\end{equation}
where $H$ is the Hamiltonian operator, $\beta=1/T$ the inverse temperature ($k_B=1$), and the trace $\rm Tr$ goes over all states 
in the Hilbert space. 
The first step of any QMC algorithm is thus the mapping of the quantum system to
an equivalent classical system, and then sample configurations of the constructed
classical systems, e.g. a system of world lines. Over the years various methods have
been developed for this mapping 
including discrete time \cite{SuzukiTrotter} and continuous time  \cite{Worms} path integrals
or the stochastic series expansion (SSE) \cite{SSE,SandvikHubbard}.
While our algorithms can be applied to any of these representations 
\cite{QuantumHistogramMethods,LectureNotes},
when combined with an extended ensemble approach
\cite{SandvikParallelTempering,SandvikQuantumParallelTempering,QWL}, 
we will concentrate on the stochastic series expansion in the following.

Most commonly, stochastic series expansion is formulated in terms of a high-temperature
series expansion of the partition function 
\begin{equation}
  Z = {\rm Tr} \exp(- \beta H) = \sum_{n=0}^{\infty} \frac{\beta^n}{n!} {\rm Tr} (-H)^n
      = \sum_{n=0}^{\infty} g(n) \beta^n \,,
  \label{Eq:SSE-ThermalExpansion}
\end{equation}
where the expansion coefficients $g(n)$ present a generalized density of states in the expansion order $n$.
This one-dimensional representation in expansion orders is similar to the classical 
form of the partition function 
\begin{equation}
  Z = \sum_E g(E) \exp(-\beta E) \,,
\end{equation}
and, in fact, we can identify low expansion orders of the thermal representation
(\ref{Eq:SSE-ThermalExpansion}) with high temperature/energy physics and higher expansion 
orders  describe low temperature/energy physics.  
In any simulation the expansion has to be truncated at some order $\Lambda$ which
is equivalent to setting a lower temperature bound. 
For canonical simulations one typically chooses $\Lambda > O(N \beta)$ such that
contributions from orders $n>\Lambda$ are negligibly small, and orders 
$n>\Lambda$ are never reached in the simulation.

We can generate a one-dimensional representation of the form in 
Eq.~(\ref{Eq:SSE-ThermalExpansion}) by direct expansion of the Hamiltonian $H$.
To this end, we decompose the Hamiltonian into a sum of $N_b$ diagonal or off-diagonal non-branching bond terms  $H = \sum_i^{N_b} H_{b_i}$ \footnote{By non-branching we mean that when applying $H_{b_i}$ to a basis state $\{\ket{\alpha}\}$, the result $H_{b_i}\{\ket{\alpha}\}$ has nonzero overlap only with at most one basis state.}.
Inserting this decomposition along with a complete set of basis states $\{\ket{\alpha}\}$ to perform the trace, the expression for the partition function becomes
\begin{equation}
  Z= \sum_{n=0}^\infty \sum_{\{\alpha\}} \sum_{\{S_n\}} \frac{\beta^n}{n!}
        \prod_{p=1}^n  \bra{\alpha(p)} (-H_{b_p}) \ket{\alpha(p-1)} \,,
  \label{Eq:SSE-BondOperatorExpansion}
\end{equation}
where  $\ket{\alpha(p)}= \prod_{i=1}^p H_{b_i} \ket{\alpha} $ denotes the propagated state 
after the action of the first $p$ bond operators on the initial state $\ket{\alpha}(=\ket{\alpha(0)}=\ket{\alpha(n)})$.
The operator string $S_n$ is an ordered sequence of $n $ bond-operators $H_{b_p}$, 
with repetitions allowed.
In order to stochastically sample the terms in the bond-operator expansion 
(\ref{Eq:SSE-BondOperatorExpansion}) by their relative weight, we need to 
interpret them as probabilities which requires all bond operators to be positive
semi-definite. For diagonal operators, we can simply achieve this by adding an energy 
offset to each bond while for off-diagonal operators there is no general 
remedy for negative weights which is the famous sign-problem for QMC calculations.

\subsection{Fixed length representation}

Most implementations of the SSE algorithm sample the various terms in the 
bond-operator expansion (\ref{Eq:SSE-BondOperatorExpansion}) by
updating the operator string $S_n$ using a fixed length representation. To this end,
a vector of length $\Lambda$ is created which contains $n$ bond operators and is
filled with $\Lambda-n$ identity operators ${\rm Id}_p$.
To compensate for the various ways to place the identity operators between the bond
operators we have to take into account a combinatorial factor ${ \Lambda \choose n}$ 
such that Eq.~(\ref{Eq:SSE-BondOperatorExpansion}) becomes
\begin{equation}
  Z = \sum_{n=0}^\Lambda \sum_{\{\alpha\}} \sum_{\{S_n\}} 
         \frac{\beta^n(\Lambda - n)!}{\Lambda!} \prod_{p=1}^\Lambda \bra{\alpha(p)} S_n(p) \ket{\alpha(p-1)} \,,
\end{equation}
where $S_n(p)$ is either a bond operator $-H_{b_p}$ of the Hamiltonian decomposition 
or an identity operator ${\rm Id}_p$.

Sampling schemes in SSE algorithms usually consist of two distinct types of updates. In a first step, attempts
are made to change the expansion order $n$ by converting diagonal bond operators to 
identities and vice versa. For the fixed length  representation, this can be achieved by
traversing the operator string $S_n$ in ascending order and at each position $p$ propose 
the update if the operator $S_n(p)$ is either a diagonal bond or identity operator. The
acceptance rates are
\begin{eqnarray}
  p_{\rm acc}\left({\rm Id}_p \rightarrow H_{b_p}\right) &=& 
     \frac{ N_d\beta \bra{\alpha(p)} H_{b_p} \ket{\alpha(p)} }{\Lambda-n}
  \nonumber \\
  p_{\rm acc}\left(H_{b_p} \rightarrow {\rm Id}_p\right) &=&
     \frac{\Lambda-n+1}{ N_d\beta \bra{\alpha(p)} H_{b_p} \ket{\alpha(p)} } \,,
  \label{Eq:SSE-FPacc}
\end{eqnarray}
where an acceptance probability $p_{\rm acc}>1$ is interpreted as $p_{\rm acc}=1$ as in 
the usual Metropolis scheme. In both expression $n$ denotes the current number of 
non-identity operators in the operator string before the update is accepted/rejected, and $N_d$ the number of diagonal bond operators
in the Hamiltonian decomposition. 

The second part of the update cycle then attempts to switch diagonal and off-diagonal 
bond operators without changing the expansion order $n$. Typically, non-local update 
techniques are employed for this step such as loop updates \cite{LoopUpdates,Henelius}, the operator loop update \cite{OperatorLoops} or 
directed loop updates \cite{DirectedLoops,GenDirLoop}.

\subsection{Variable length representation}
\label{Sec:VariableLength}

Alternatively to the fixed length representation outlined above we can keep the original representation of an operator string with a variable number $n$ of (non-identity) bond operators.
Since this list representation of the operator string turns out to be the computationally 
more efficient representation in a broad-histogram simulation, we will discuss in the following an
efficient
sequential update algorithm which inserts or removes diagonal bond operators from the 
operator string for this representation, extending on previous work
by D.~C. Handscomb~\cite{Handscomb} and A.~W. Sandvik~\cite{SandvikHubbard}.
 
\begin{figure}[t]
  \includegraphics[width=\columnwidth]{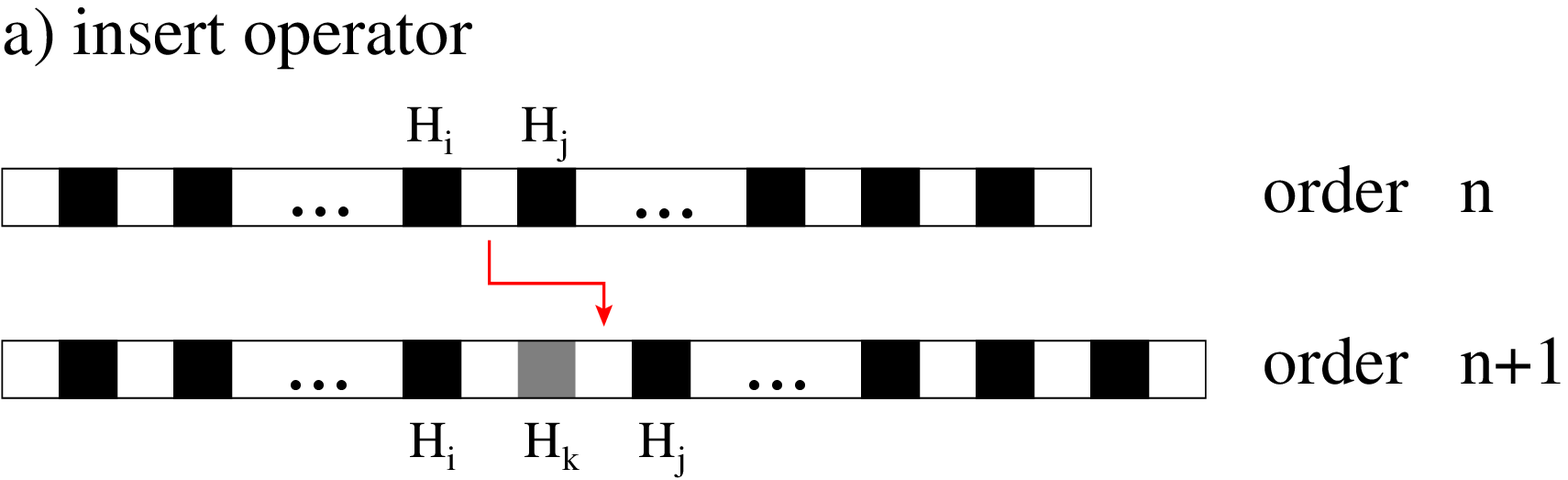} 
  \vskip 5mm
  \includegraphics[width=\columnwidth]{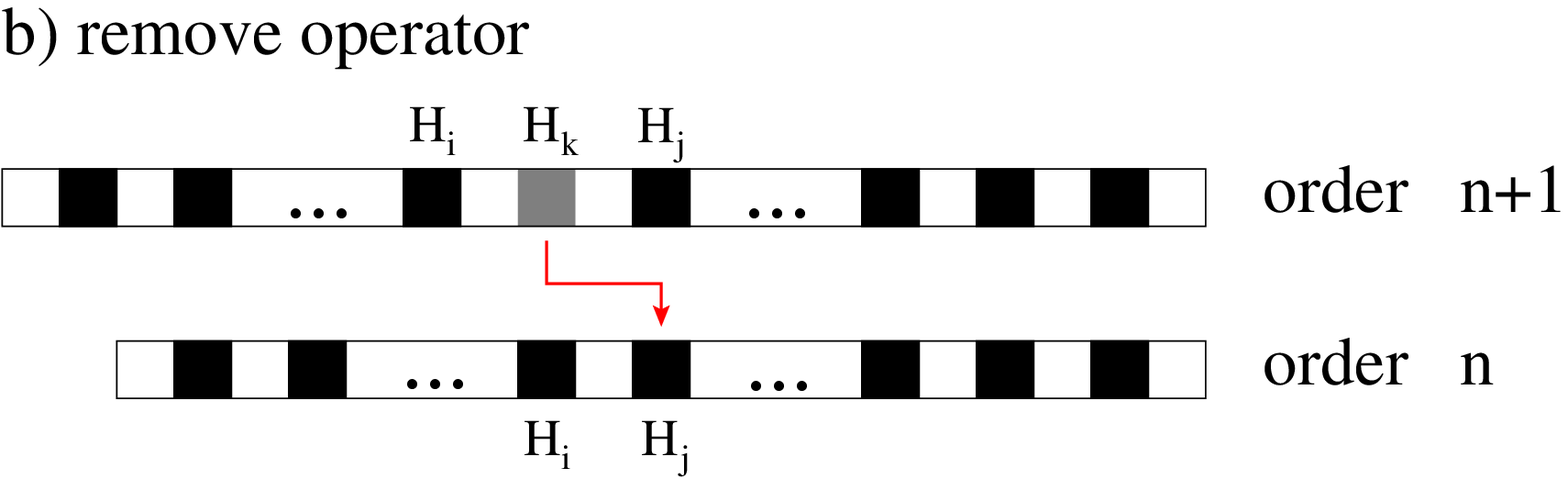}
  \caption{
  Variable length representation of the operator string in a stochastic series 
  expansion: Bond operators from the Hamiltonian decomposition are marked
  by filled squares. Between each pair of bond operators an empty square denotes
  the ``gap" which in an update move can be filled with an additional diagonal bond 
  operator as shown in the insert move in the top panel. 
  In the reverse move shown in the lower panel a diagonal bond operator is 
  removed from the operator string. 
  The arrow indicates for which position of the operator string a subsequent
  update will be proposed.
  }
  \label{Fig:InsertRemoval}
\end{figure}

We start with an operator string as illustrated in Fig.~\ref{Fig:InsertRemoval}, where 
non-identity bond operators from the Hamiltonian decomposition are depicted by filled 
squares. These are separated by ``gaps'' into which additional operators can be inserted.
Starting from the gap preceding the lowest-order bond operator we traverse the operator
string in ascending order, alternatingly updating ``gaps'' and operators:
\\

\begin{itemize}
\item When updating a gap we try to insert a randomly chosen
diagonal bond operator, and accept the insertion with probability
\begin{equation}\label{insprob}
 p_{\rm insert} = \min\left[1,
     \frac{ N_d\beta \bra{\alpha(p)} H_{b_p} \ket{\alpha(p)} }{n+1}\right],
\end{equation}
where $n$ denotes the size of the operator string before the update.

If the insertion succeeds we continue with attempts to insert additional operators after the one just inserted, as illustrated in the top panel of Fig.~\ref{Fig:InsertRemoval}, and continue inserting as long as the insertions are accepted. At the first rejected insertion we stop updating this gap and continue updating the following operator.
\item When updating a bond operator we attempt to remove the current operator if it is diagonal, with probability
\begin{equation}\label{remprob}
  p_{\rm remove}=\min\left[1,
     \frac{n}{ N_d\beta \bra{\alpha(p)} H_{b_p} \ket{\alpha(p)} } \right]\,,
\end{equation}
where again $n$ denotes the size of the operator string before the update.

If the removal succeeds we next attempt removing the following bond operator and continue removing bond operators until an attempt is rejected. We then continue updating the following gap.
\end{itemize}

The subsequent loop update in the SSE sampling scheme which swaps diagonal and
off-diagonal bond operators without changing the expansion order remains unchanged
for the variable length representation. 
This algorithm fulfills detailed balance as shown in appendix \ref{Sec:Proof}.

Detailed balance is fulfilled not at the level of individual update steps, but only at the level of complete sweeps. Hence measurements of physical observables have to be done only after a full sweep
of updates, i.e. a complete traversal of the operator string. 

\section{Optimized ensembles}
\label{Sec:ExEn}

The stochastic series expansion is typically performed for a canonical ensemble 
at a fixed inverse temperature $\beta$. From the high-temperature expansion of 
the partition function in Eq.~(\ref{Eq:SSE-ThermalExpansion}) it is then obvious that 
the simulated random walk will concentrate on a narrow energy window, or a small 
number of expansion orders characteristic for this energy regime respectively
\begin{equation}\label{eq:EnergyEstimator}
E=\langle H \rangle = - \frac{\partial}{\partial \beta} \ln Z = - \frac{1}{Z} \sum_{n=0}^\infty g(n) n \beta^{n-1} = - \frac{1}{\beta} \langle n \rangle \,.
\end{equation}
For a system with competing energy scales which are typically found in systems undergoing
a first-order transition or systems with rough energy landscapes the canonical sampling
can result in dramatic slowing-downs with the random walker stuck in one of the energy levels
and the tunneling to another energy level suppressed by some intermediate energy barrier.
To overcome this problem parallel tempering or broad-histogram algorithms aim at
broadening the sampled energy space by introducing either multiple replicas of the system
spread over some temperature range or by introducing an extended ensemble that creates
an additional bias for the random walker to tunnel through the intermediate energy barriers. 
In the following we  concentrate on the idea of generalizing the sampled statistical ensemble
in the context of a stochastic series expansion. 

A first step in this direction has been taken by formulating the iterative algorithm by 
Wang and Landau \cite{WangLandau} to sample a broad range of expansion orders in
the SSE algorithm \cite{QWL}. The goal of the algorithm is to approach a flat histogram
ensemble that samples all expansion orders (up to some cutoff $\Lambda$) equally
often. Similar to the classical case such uniform sampling is achieved by introducing
an additional statistical weight $w(n)$ that is inversely proportional to a generalized 
density of states $g(n)$, that is 
\begin{eqnarray}
 w_{\rm flat}(n) \propto \frac{1}{g(n)} && {\rm fixed\,\, length\,\, representation}, \nonumber \\
 w_{\rm flat}(n) \propto \frac{2n+1}{g(n)} && {\rm variable\,\, length\,\, representation}, \quad
 \label{Eq:FlatWeights}
\end{eqnarray}
where the additional factor of $2n+1$ is due to the $2n+1$ local steps in one sweep through
the operator string in the variable length representation (see Sec. \ref{Sec:VariableLength})
in contrast to the fixed length representation where the effort per sweep is constant.
Accordingly the acceptance rates are multiplied by this additional weight factor and 
setting the inverse temperature to $\beta=1$. For the fixed-length representation of 
the operator string the acceptance probabilities in Eq.~(\ref{Eq:SSE-FPacc}) become
\begin{eqnarray}
  p_{\rm acc}\left({\rm Id}_p \rightarrow H_{b_p}\right) &=& 
     \frac{w(n+1)}{w(n)} \cdot \frac{ N_d \bra{\alpha(p)} H_{b_p} \ket{\alpha(p)} }{\Lambda-n} \quad
  \nonumber \\
  p_{\rm acc}\left(H_{b_p} \rightarrow {\rm Id}_p\right) &=&
     \frac{w(n-1)}{w(n)} \cdot \frac{\Lambda-n+1}{ N_d \bra{\alpha(p)} H_{b_p} \ket{\alpha(p)} } \quad
  \label{Eq:SSE-OFPacc}
\end{eqnarray}
and for the variable-length representation of the operator string we obtain
\begin{eqnarray}
  p_{\rm acc}\left(\square_p \rightarrow H_{b_p}\right) &=& 
     \frac{w(n+1)}{w(n)} \cdot \frac{ N_d\bra{\alpha(p)} H_{b_p} \ket{\alpha(p)} }{n+1}
  \nonumber \\
  p_{\rm acc}\left(H_{b_p} \rightarrow \square_p\right) &=&
     \frac{w(n-1)}{w(n)} \cdot \frac{n}{ N_d \bra{\alpha(p)} H_{b_p} \ket{\alpha(p)} } \quad \;\;
  \label{Eq:SSE-OVPacc}
\end{eqnarray}
In the following we will further expand on this idea and ask which weights $w(n)$ are
optimal for a given physical system in the sense that the sampled statistical ensemble 
gives fastest equilibration for all thermodynamic observables.
For classical systems it has been demonstrated that such optimal weights exist and
generally lead to a non-uniform histogram \cite{OptimizedEnsembles}. Typically, the histogram
is reweighed and additional resources (in terms of attempted updates) are shifted towards
those values of the reaction coordinate (in which the statistical ensemble/weights are
defined) where bottlenecks of the simulation occur, typically in the vicinity of a phase transition. 
The optimal weights are systematically approached in Ref.~\onlinecite{OptimizedEnsembles}
by studying a diffusive current of random walkers between the extremal values of the reaction 
coordinate and feeding back measurements of the local diffusivity. 

In complete analogy, we can think of the random sampling of operator string configurations 
during a SSE simulation as a one-dimensional random walk in expansion orders. 
To maximize equilibration we want this random walker to perform as many round-trips 
between the two extremal expansion orders $n_{\rm min}=0$ and $n_{\rm max}=\Lambda$
as possible. 
To identify such round trips we  add a label to the random walker that keeps track of
which of the two extremal orders the random walker has visited most recently, e.g. ``+" and 
``-" denote that the walker has visited the lowest expansion order $n_{\rm min}$ or the 
highest expansion order $n_{\rm max}$ most recently. 
The label is switched only once the random walker visits the other extremal order. 
Using this label we can record two histograms $h_+(n)$ and $h_-(n)$ where after every 
single update of the operator string we increment the histogram corresponding to the current label
of the walker.
This allows us to measure for each expansion order what fraction of walkers 
$f(n) = h_+(n) / (h_+(n) + h_-(n))$ on average are passing by a given expansion order 
that have visited, for instance, the lowest expansion order last. 
We can then define to leading order a current that characterizes the diffusion of the random
walker from the lowest to the highest expansion order
\begin{equation}
  j = D(n) h(n) \frac{df}{dn} \,, 
  \label{Eq:Current}
\end{equation}
where $D(n)$ is the local diffusivity, and $h(n)=h_+(n) + h_-(n))$ the  local histogram of visits 
of the random walker to the
expansion order $n$.
In order to speed up equilibration we want to maximize this diffusive current by varying the 
sampled histogram $h(n)$.
Following the arguments in Ref.~\onlinecite{OptimizedEnsembles} the optimal histogram of visits
is  found to be inversely proportional to the square root of the local diffusivity $D(n)$
\begin{equation}
  h^{\rm opt}(n) \propto 1/\sqrt{D(n)} \,,
  \label{Eq:OptimalHistogram}
\end{equation} 
where the local diffusivity can be estimated from the current histogram of visits by
\begin{equation}\label{eq:diffesti}
  D(n) \propto \frac{1}{h(n) \cdot |df/dn|} \;.
\end{equation}
For the example  systems discussed in section \ref{Sec:Examples} we
find -- similar to simulations with classical systems -- that this local diffusivity depends
only weakly on the sampled statistical ensemble. The sampled histogram, however,
reflects a dramatic reweighing of computational resources towards those expansion 
orders where the local diffusivity is suppressed, typically those orders which can be 
identified with a thermal or quantum phase transition as discussed in section 
\ref{Sec:Examples}.

\subsection{Feedback algorithm}

In order to obtain a statistical ensemble with weights $w^{\rm opt}(n)$ which will
produce the optimal broad histogram in Eq.~(\ref{Eq:OptimalHistogram}) we can 
apply iterations of the feedback algorithm outlined in Ref.~\onlinecite{OptimizedEnsembles}.
The algorithm starts from an arbitrary broad-histogram ensemble, typically a 
flat-histogram ensemble with approximative weights  obtained from a few
iterations of the quantum Wang-Landau algorithm \cite{QWL}. 
The optimized ensemble is then systematically approached by feeding back the 
local diffusivity $D(n)$ estimated for the current set of statistical weights. To ensure
convergence in subsequent feedback iterations, statistical measurements need
to be improved which we accomplish by increasing the number of simulated
update sweeps for each iteration. 
The algorithm can be outlined as follows:
\begin{itemize}
\item Start with some trial weights $w(n)$,\\ such as $w(n) \approx  1/g(n)$.
\item Repeat
\begin{itemize}
\item Reset the histograms\\ $h(n) = h_+(n) = h_-(n) = 0$.
\item Simulate the system for $N_{\rm sw}$ sweeps:
\begin{itemize}
\item When traversing the operator string updates\\ are accepted
          according to Eqs. (\ref{Eq:SSE-OFPacc},\ref{Eq:SSE-OVPacc}).
\item After each step of the traversal update\\ 
          the labeled histograms $h_+(n)$ and $h_-(n)$\\
          and the walker's label.
\item Perform a series of loop updates without\\ changing the expansion order. 
\item Update the global histogram $H(n)$\\
          at the end of the sweep.
\end{itemize}
\item Calculate the fraction $f(n) = \frac{h_+(n)}{h_+(n) + h_-(n)}$
\item Define new statistical weights as
\[
   w(n) \leftarrow w(n) \cdot \sqrt{\frac{1}{h_+(n)+h_-(n)} \cdot \left|\frac{df}{dn}\right| } \;.
\]
          and recalculate the acceptance probabilities.
\item Increase the number of sweeps $N_{\rm sw} \leftarrow 2N_{\rm sw}$.
\end{itemize}
\item Stop once the histogram $h(n)$ has converged.
\end{itemize}

Note  that the local histograms $h_+(n)$ and $h_-(n)$ are updated after every
attempted move when traversing the operator string to update the sequence of diagonal
bond operators. 

For measurements we need to additionally measure a global histogram $H(n)$  after each full sweep. 
This global histogram $H(n)$ is in particular employed to
estimate the expansion coefficients $g(n)$ from 
\begin{equation}\label{eq:getexpcoeff}
g(n)\propto H(n)/w(n),
\end{equation}
in terms of the final statistical weights, and normalized via $g(0)=d_H$, the dimensionality of the Hilbert space.

In a fixed length representation, the expectation values of the histograms (prior to normalization) are related as $h(n)=\Lambda \cdot H(n)$ and there is thus no difference after normalization.
In the variable length representation, the average number of operator insertion/removal attempts scales with the initial 
expansion order as $2n+1$, so that the  number of visits $h(n) \approx (2n+1)\cdot H(n)$. 

Based on the values of the expansion coefficients $g(n)$, obtained using Eq.~(\ref{eq:getexpcoeff}), we can  calculate thermodynamic properties of the system. In particular, the free energy at an inverse temperature $\beta$ is obtained as
\begin{equation}\label{eq:FreeEnergyEstimator}
F=-\frac{1}{\beta}\ln{Z}=-\frac{1}{\beta}\ln \sum_n g(n)\beta^n,
\end{equation}
the energy from Eq.~(\ref{eq:EnergyEstimator}) as
\begin{equation}
E=\frac{1}{Z}\sum_n n \: g(n) \beta^{n-1},
\end{equation}
and the entropy using $S=(E-F)/T$.
In order to obtain thermal expectation values of an observable $A$ (such as e.g. the magnetization), one in addition records separately the values of the observable $A$ at each expansion order $n$, taking measurements after each full Monte Carlo step: Denoting by $A_i$  the values of the observable $A$ from its measurement performed after the $i$-th full Monte Carlo step, and by $n_i$ the value of the expansion order of the corresponding SSE configuration, the mean values  of the observable $A$ at expansion order $n$ is given in terms of the global histogram $H(n)$ as
\begin{equation}
A(n)=\frac{1}{H(n)}\sum_{i, n_i=n}  A_i,
\end{equation}
where the summation is restricted to those values of $i$, for which $n_i=n$.
From these mean values $A(n)$ of the observable $A$ for each expansion order, one  obtains the thermal expectation value at an inverse temperature $\beta$ as
\begin{equation}
\langle A \rangle = \frac{1}{Z}{\rm Tr} \:A \exp(- \beta H) =\frac{1}{Z}\sum_n A(n) \: g(n) \beta^n.
\end{equation}
Since only a finite number of expansion orders (up to the cutoff $\Lambda$) can be treated within the QMC simulations, there  exists  
a limited temperature range, over which both the thermodynamic quantities and  observables can be calculated, while a pronounced runaway has been observed beyond this temperature regime~\cite{QWL}.
Furthermore, statistical errors  are reliably estimated from repeating the full procedure using independent random number sequences.

In our application of the feedback algorithm to a variety of quantum spin Hamiltonians
we have found that a few iterations, typically four, are sufficient to obtain convergence
of the sampled histograms. The initial trial weights are generated by running a few
iterations of the quantum Wang-Landau algorithm. For the purpose of the
subsequent ensemble optimization it is not necessary for the initial weights to be well
converged to the flat-histogram ensemble, e.g. $w(n) = 1/g(n)$, but should simply allow
to sample a broad histogram over all expansion orders such that round trips between
the extremal expansion orders occur at a sufficiently high rate such that the sampled
histogram $H(n)$ and the derivative of the fraction $f(n)$ needed for the feedback can 
be estimated.

\input{examples.tex}

\section{Conclusions}
We presented an application of the optimized broad-histogram ensemble method to the simulation of quantum systems, 
in order to increase the performance of previously introduced extended ensemble methods.
We found that within the stochastic series expansion approach, the fixed length operator string representation suffers from an insufficient local simulation dynamics at the end of the range for small and large expansion orders $n$, which can be overcome by formulating the algorithm using a variable length operator string. We derived the appropriate update probabilities for this scheme. We then adapted a recently developed feedback algorithm to the quantum case, and provided a performance analysis of the resulting algorithm.

The analysis of the optimized ensemble approach in the variable length representation for the Heisenberg chain and the second-order thermal phase transition in the Heisenberg model on a cubic lattice,
suggests that for quantum models showing no, or  second-order thermal phase transitions, the optimal ensemble is characterized by a {\it flat histogram} $h(n)$ if a {\it variable length} formulation of the SSE is used. 
This flat histogram $h(n)$ counting local updates corresponds to a histogram $H(n)\propto 1/(2n+1)$ at the level of full sweeps. 
In such computationally "easy" cases it is thus feasible to directly 
incorporate the optimal ensemble weights 
of Eq.~(\ref{Eq:FlatWeights}) 
in the quantum Wang-Landau approach, using the variable length representation Eq.~(\ref{Eq:SSE-OVPacc}).

In our analysis of the  first-order thermal transition of interacting hard-core bosons and the spinflop transition, we found that the optimized ensemble approach
shifts the distribution of resources towards the transition region, thus allowing for a faster equilibration as compared to flat-histogram methods, although the scaling with system size remains exponential for strongly first order transitions.
This suggests that  broad-histogram ensemble optimization in a single reaction coordinate might not 
provide a general remedy of such exponential slowing-down.

\paragraph*{Acknowledgments}
We thank N. Hodler for his participation in the initial stages of this project
and F. Alet and A.~W. Sandvik for stimulating discussions. We acknowledge hospitality of the 
Kavli Institute for Theoretical Physics, Santa Barbara, and the Aspen Center for Physics.
S. W. acknowledges HLRS Stuttgart and NIC J\"ulich for allocation of computing time.
Some of our numerical simulations were based on the ALPS libraries \cite{ALPS}.
This research was supported in part by the Swiss National Science Foundation, the German Research Foundation,
and the National Science Foundation under Grand No. NSF PHYS05-51164.

\appendix

\section{Proof of detailed balance for the variable length operator string representation}
\label{Sec:Proof}
\input{appendix.tex}

\end{document}

%% file: examples.tex
\section{Performance Studies}
\label{Sec:Examples}
We next present results from a systematic performance analysis of the new algorithm for various quantum lattice models, including a comparison to flat-histogram techniques.
In the following, we label results obtained using a flat-histogram ensemble by "flat-fixed" and "flat-var" for the fixed
and variable length operator string SSE representation, respectively. Results obtained from the optimized ensemble are denoted as 
"opti-fixed" and "opti-var"
for the fixed
and variable length representation, respectively.

\subsection{The spin-1/2 Heisenberg chain}
We start by considering the case of a spin-1/2 Heisenberg chain. This allows us 
to compare the  various methods first in the absence of any phase transition.
In later sections, we then focus on the additional issues arising from the presence of  first- and second-order phase transitions.

We simulate the spin-1/2 Heisenberg model on a finite chain of $N=10$ sites with periodic boundary conditions, described by the 
Hamiltonian
\begin{equation}
H=J \sum_{i} \mathbf{S}_i \cdot \mathbf{S}_{i+1},
\end{equation}
where $J>0$, and $\mathbf{S}_i $ denotes a spin-1/2 degree of freedom at site $i$.
For this system, we calculated the exact thermal expansion coefficients $g_{\rm exact}(n)$ up to order $n=\Lambda=500$, based on a full numerical diagonalization of the Hamiltonian. We used these values in order to perform fixed-weights simulations in a flat-histogram ensemble.
For the fixed length representation, we set the weights $w(n)=1/g_{\rm exact}(n)$; and for the variable length representation, we use $w(n)=(2n+1)/g_{\rm exact}(n)$.
In both cases we ran 16 parallel simulations with independent random number streams in order to assess the statistical variations due to the QMC sampling. 
We furthermore performed up to 4 feedback iteration batches to obtain  converged optimized ensembles, where the number of MC steps was increased by a factor of 2 after each feedback step. Again, this procedure was repeated with 16 independent random number sequences. 
In all cases, we employed the multi-cluster SSE loop update~\cite{Henelius,NonLocalUpdates}, flipping 
each 
cluster 
that results from the deterministic operator loop construction with probability 1/2.

\begin{figure}[t]
  \includegraphics[width=\columnwidth]{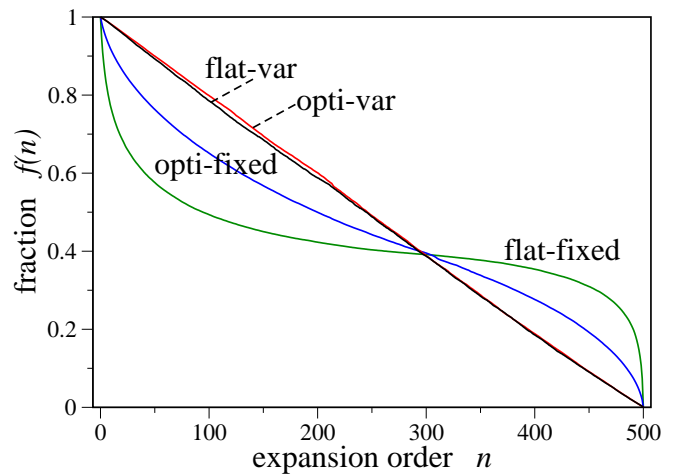}
  \caption{
  Average fraction $f(n)$ of random walkers diffusing from lowest order
  (high temperatures) to highest orders (low temperatures) in a stochastic
  series expansion in the inverse temperature of a spin-1/2 Heisenberg
  chain with 10 sites.
  Data is shown for the optimized ensemble with fixed length operator
  string (opti-fixed) and variable length operator string (opti-var), as well as data
  for the flat histogram ensemble for both operator string representations
  (flat-fixed and flat-var).
  }
  \label{Fig:ChainFraction}
\end{figure}
\begin{figure} [t]
  \includegraphics[width=\columnwidth]{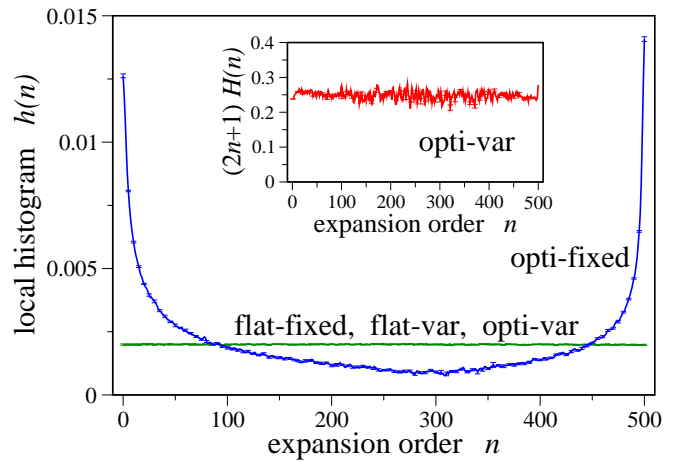}
  \caption{
  Sampled local histograms $h(n)$ for the optimized ensemble using a
  variable length operator string (opti-var) and fixed length operator
  string (opti-fixed), as well as the corresponding flat-histogram methods (flat-fixed and flat-var),  
  in a stochastic series expansion in the inverse temperature 
  of a spin-1/2 Heisenberg chain with 10 sites.
  The inset shows the flat  rescaled global histogram $(2n+1)\cdot H(n)$ for the 
  optimized ensemble with variable length operator string.
  }
  \label{Fig:ChainHistogram}
\end{figure}

In Fig.~\ref{Fig:ChainFraction}, the averaged fractions $f(n)$ are shown for the various algorithms, and Fig.~\ref{Fig:ChainHistogram} provides the corresponding local histograms.
Let us first consider the flat-fixed method, which is based on a flat histogram ensemble in the fixed length representation. We find that for this method, the 
fraction $f(n)$ in Fig.~\ref{Fig:ChainFraction} exhibits steep descents near both ends, $n=0$ and $n=\Lambda$, of the $n$-range.  As seen from Eq.~(\ref{eq:diffesti}),
the increased derivative $df/dn$ indicates a suppressed diffusivity of the random walker near both ends of the $n$-range.
This slowing-down near $n=0$ and $n=\Lambda$  is however not related to physical properties of the system under consideration.
Instead, it is due to an inefficient local bond-operator insertion/removal dynamics in the fixed length operator string representation. This is seen from
the acceptance probabilities in Eqs.~(\ref{Eq:SSE-OFPacc}): for small $n\ll \Lambda$ the insertion process is suppressed by a large denominator, and for $n$ close to $\Lambda$, the removal process is suppressed by a small numerator. 
The ensemble optimization tries to compensate for this technical inefficiency by shifting resources towards both ends of the $n$-range, 
as seen from the histogram shown in Fig.~\ref{Fig:ChainHistogram}. However, this allocation of resources near \textit{both} ends of the $n$-range is not efficient, as the computational effort for performing updates scales proportional to the expansion order $n$. It would thus be desirable to overcome the slowing-down of the fixed length operator string representation by an improved 
bond-operator insertion/removal dynamics. 
This motivated us to consider the  variable length operator string SSE representation introduced in Sec. II.

Indeed, 
changing to the flat-histogram ensembe in the variable length representation, 
by using the flat-var method, the slowing-down is completely eliminated, as seen from Fig.~\ref{Fig:ChainFraction}, which shows an almost linearly decreasing fraction of the flat-var method over the whole $n$-range.
Fig.~\ref{Fig:ChainFraction} shows that the optimized ensemble method opti-var leads to a very similar 
linear  fraction $f(n)=1-n/\Lambda$. Indeed, both the flat-var and the opti-var method show a flat local histogram $h(n)$, seen in 
in Fig.~\ref{Fig:ChainHistogram}.
Interestingly, we thus find that for the Heisenberg chain, the flat-var method already provides an optimal ensemble, in the sense that also the opti-var method leads to a flat local histogram $h(n)$.
This corresponds in both cases to a uniform diffusivity of the random walker throughout the whole $n$-range.

Using the variable length representation, the simulation of the Heisenberg chain  does not suffer anymore from any slowing-down, in accordance with the absence of phase transitions in this model. 
The global histogram $H(n) \approx h(n)/(2n+1)$ then decreases with $n$ like $H(n)\propto 1/(2n+1)$. This is indeed seen from the inset of Fig.~\ref{Fig:ChainHistogram}.
Since the computational effort of the SSE update methods scales linear with $n$, this implies that an equal amount of resources is devoted to each expansion order within the considered $n$-range.

\begin{table}
\begin{tabular}{|l|l|l|l|l|}
\hline
Method & $\tau_{\rm up}/10^6$ & $\tau_{\rm down}/10^6$ & $\tau_{\rm round}/10^6$ & $ \tau_{\rm round}\: / \: \tau^{opti-var}_{\rm round}$\\
\hline
flat-fixed & 1.21(1) & 1.62(1) & 2.83(2) & 6.86(5)\\
opti-fixed & 0.892(3) & 0.880(3) & 1.772(5) & 4.30(1)\\
flat-var & 0.202(3) & 0.210(3) & 0.412(3) & 1 \\
opti-var & 0.207(1) & 0.205(2) & 0.412(3) & 1\\
\hline
\end{tabular}
\label{Tab:ChainTimes}
\caption{Traversal times of random walkers diffusing from lowest order
  to highest orders in a stochastic
  series expansion in the inverse temperature of a spin-1/2 Heisenberg
  chain with 10 sites.
  Data is shown for the optimized ensemble with fixed length operator
  string (opti-fixed) and variable length operator string (opti-var), and the flat histogram ensemble for both operator string representations
  (flat-fixed and flat-var) in units of $10^6$ operator insertion/removal attempts. The statistical error, estimated by performing 16 independent runs in each case is given for the least significant digit.}
\end{table}

The dynamics of the simulation in sweeping through the $n$-range can be quantified in terms of the traversal times of the random walker. For this purpose, we measure the averaged traversal time $\tau_{\rm up}$ ($\tau_{\rm down}$) of the random walker to travel from the lowest (highest) to the highest (lowest) expansion order, in units of single operator insertion/removal attempts.
 Furthermore, we denote by $\tau_{\rm round}=\tau_{\rm up}+\tau_{\rm down}$ the round-trip time of the random walker.
In Tab.~\ref{Tab:ChainTimes} the  traversal times are compared for the various ensembles and representations, averaged over the 16 independent runs in each case. 
For  the fixed length representation, we find that in the optimized ensemble 
the random walker's round-trip time $\tau_{\rm round}$ is significantly reduced, as compared to the corresponding flat histogram ensemble. 
Furthermore, for the optimized ensembles, $\tau_{\rm up}$ and $\tau_{\rm down}$ are similar, meaning that the random walker spends about the same number of insertion/removal attempts in both directions, whereas in the flat histogram ensembles, $\tau_{\rm down}$ is larger than $\tau_{\rm up}$. 
For
the opti-fixed approach we obtained a larger round-trip time $\tau_{\rm round}$ than for both the flat-var and the opti-var method.
This shows that the improved dynamics of the variable length representation provides a performance improvement 
that cannot even be achieved by the optimized ensemble using the fixed length approach. That the flat-var method is already close to optimal,
is  reflected by the equal traversal times for the flat-var and the opti-var method in Tab.~\ref{Tab:ChainTimes}.

We next assess the extend to which the  improvements in the variable length representation reduces the statistical error of the expansion coefficients $g(n)$,
which are estimated using Eq.~(\ref{eq:getexpcoeff}) from the global histogram $H(n)$ and the employed weights $w(n)$, 
normalized via $g(0)=d_H$, the dimension of the Hilbert space (here, $d_H=2^{N_s}$).  To quantify the error, we calculated the logarithmic deviation 
\begin{equation}
\delta \ln [g(n)]=\ln\left[g(n)\right]-\ln\left[g_{ex}(n)\right]=\ln\left[g(n)/g_{ex}(n)\right]
\end{equation}
to the exact expansion coefficients, after running each simulation for the same,  fixed CPU time. 
\begin{figure}
  \includegraphics[width=\columnwidth]{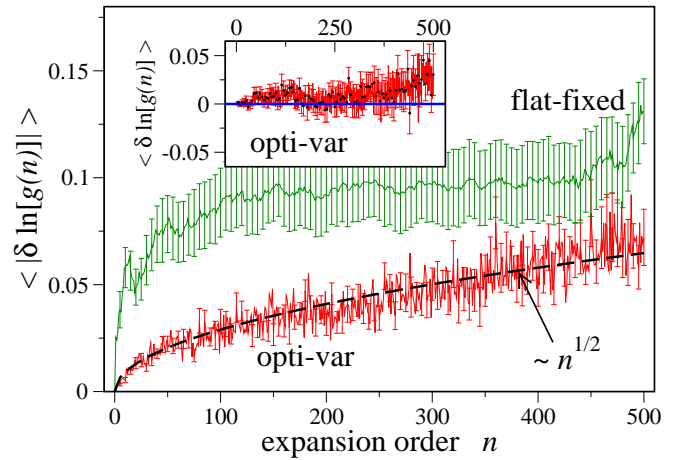}
  \caption{
  Averaged absolute deviation of the calculated expansion coefficients $g(n)$
  from the exact result for a spin-1/2 Heisenberg chain with 10 sites. 
  Data for the flat histogram ensemble with fixed length operator string (flat-fixed) 
  and the optimized ensemble with variable length operator string (opti-var) are shown,
  both averaged over 16 independent runs.
  The inset shows the fluctuation around the exact result (zero line) of the averaged statistical deviation for the estimate
  calculated in the optimized ensemble. 
  }
  \label{Fig:ChainErrorDOS}
\end{figure}
In Fig.~\ref{Fig:ChainErrorDOS}
the absolute deviation, averaged over 16 independent runs, $\langle | \delta \ln [g(n)] | \rangle$ is shown vs. $n$ for the opti-var 
(corresponding in the current case to the flat-var method) and the flat-fixed method.  We obtain a significant overall improvement  using 
the opti-var ensemble, in particular at low values of $n$.
The inset of Fig.~\ref{Fig:ChainErrorDOS}, showing the averaged deviation $\langle \delta \ln [g(n)]\rangle $ for the opti-var method, verifies the agreement within statistical errors of the QMC results with the exact expansion coefficients. 

We find the averaged absolute deviation to scale $\langle | \delta \ln [g(n)] | \rangle\propto n^{1/2}$ for the opti-var method. 
This relates to the approximate
$H(n)\propto 1/n$
scaling of the  
global histogram, as expected from the standard scaling of the MC error with the number of samples:
sampling the systems at an expansion  order $n$ a number of $H(n)\propto 1/n$ times, the statistical error in $g(n)\propto H(n)/w(n)$ is  antici\-pated to scale 
proportional to 
$1/\sqrt{H(n)}\propto \sqrt{n}$. 

From the expansion coefficients one can calculate thermodynamic 
quantities of the model down to  temperatures $T_{min}$, that scale as $T_{min}\propto 1/\Lambda$ with the cutoff $\Lambda$~\cite{QWL}. Here, we consider temperatures $T$ down to $0.03J$, below which strong deviations result due to the finite $n$-range~\cite{QWL}.
Based on the fixed CPU-time QMC results for $g(n)$, we calculated the free energy $F$ from Eq.~(\ref{eq:FreeEnergyEstimator}), and 
its deviation 
$\delta F=F-F_{ex}$
to the exact free energy $F_{ex}$, obtained from  the $g_{ex}(n)$ in Eq.~(\ref{eq:FreeEnergyEstimator}) instead.
\begin{figure} 
  \includegraphics[width=\columnwidth]{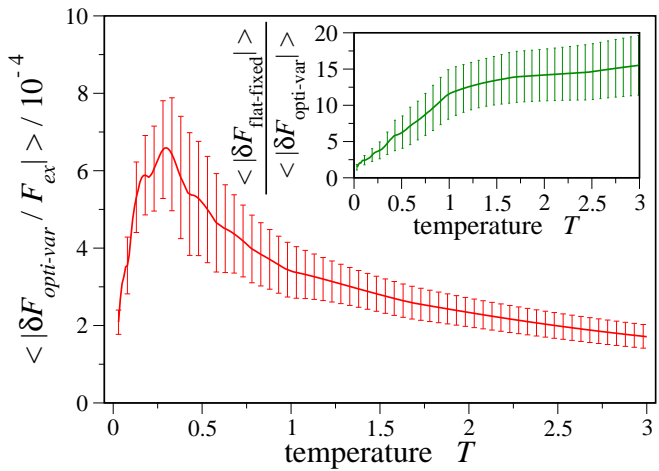}
  \caption{
  Averaged absolute values of the relative deviation of the calculate free energy $F$ 
  for the optimized ensemble using a variable length operator string (opti-var)
  from the exact result for a spin-1/2 Heisenberg chain with 10 sites. 
  The data is averaged over 16 independent runs.
  The inset shows the averaged relative  deviation of the estimated free energy
  from flat histogram simulations using fixed length operator
  strings (flat-fixed), compared to the optimized ensemble method (opti-var). 
  }
  \label{Fig:ChainErrorFreeEnergy}
\end{figure}
Figure~\ref{Fig:ChainErrorFreeEnergy} shows the averaged absolute values of the relative deviation $\langle |\delta F/F_{ex}|\rangle$ for the opti-var method. 
Compared to the flat-fixed method, 
the opti-var approach shows significantly reduced errors in the free energy over the whole temperature range, 
in particular reducing the deviation by an order of magnitude at high temperatures, $T>J$ (see the inset of 
Fig.~\ref{Fig:ChainErrorFreeEnergy}).


\subsection{Second-order phase transition}\label{Sec:Cube}
In our analysis of the spin-1/2 Heisenberg chain, we found that the variable length representation eliminates the technical problems with the fixed length representation. Already the flat-var approach produces an optimal ensemble with no features in the diffusivity, which would otherwise indicate a slowing-down of the simulation dynamics in certain regions of the $n$-range. 
This behavior is expected, as the Heisenberg chain  does not exhibit any finite temperature phase transition, and the multi-cluster updates provide an efficient sampling scheme.

We now consider a system that does exhibit a second-order thermal phase transition. In particular, we simulate the spin-1/2 Heisenberg model 
\begin{equation}
H=J \sum_{<i,j>} \mathbf{S}_i \cdot \mathbf{S}_{j},
\end{equation}
on a three-dimensional cubic lattice, with an antiferromagnetic nearest neighbor coupling $J>0$ and periodic boundary conditions. 
Here, we focus on a cube of $N_s=8^3$ sites. In the thermodynamic limit, the model exhibits a 
second-order phase transition at a temperature $T_c\approx 0.946~$\cite{SandvikParallelTempering}, which 
corresponds on this finite system to an expansion order of around $n_c \approx 900$.
Since no exact results for $g_{ex}(n)$ are accessible for this system, we
first performed 8 Wang-Landau sampling steps starting from a uniform assignment $g(n)=1, n=0,...,\Lambda=4000$, in order to obtain a broad histogram ensemble. Then, we performed 4 feedback iteration batches, where the number of MC steps was increased by a factor of 2 after each feedback step. Similar to the previous case, we performed 16 independent simulations for statistical averaging  the final results. Using the final estimates of the 
expansion coefficients $g(n)$, we then performed simulations in the flat-histogram ensemble for comparison to the optimized ensembles. 
We generally employed multi-cluster updates, except for the
opti-var method, for which also single cluster (s.c.) updates were considered. 
In each  single cluster update step, only one of the clusters resulting from the deterministic loop 
construction 
is flipped.
This was done in order to artificially reduce the efficiency of the loop update scheme, and allows us to study the optimized ensemble when using a less efficient SSE loop update scheme with increased autocorrelations near the critical temperature.

\begin{table}
\begin{tabular}{|l|l|l|l|l|}
\hline
Method & $\tau_{\rm up}/10^7$ & $\tau_{\rm down}/10^7$ & $\tau_{\rm round}/10^7$ & $ \tau_{\rm round}\: / \: \tau^{opti-var}_{\rm round}$\\
\hline
flat-fixed & 10(1)& 7.8(5) & 18(1) & 7.2(5) \\
opti-fixed & 5.87(5)& 5.72(5) & 11.59(7)& 4.6(1)\\
flat-var &  1.08(2)& 1.39(3) & 2.46(2) &  1\\
opti-var & 1.252(5)& 1.246(4)& 2.47(6)  & 1\\
opti-var s.c.& 3.13(5) &3.12(5) &6.25(5) &2.5(1)\\
\hline
\end{tabular}
\label{Tab:CubeTimes}
\caption{Traversal times of random walkers diffusing from lowest order
  to highest orders in a stochastic
  series expansion in the inverse temperature of a spin-1/2 Heisenberg
  cube with $8^3$ sites.
  Data is shown for the optimized ensemble with fixed length operator
  string (opti-fixed) and variable length operator string (opti-var), as well as data
  for the flat histogram ensemble for both operator string representations
  (flat-fixed and flat-var) in units of $10^7$ operator insertion/removal attempts. We employed multi-cluster updates, except for the
opti-var method, for which using only single cluster (s.c.) updates was considered.
  The statistical error, estimated by performing 16 independent runs in each case is given for the least significant digit.}
\end{table}

The resulting traversal times for the various methods are given in  Tab.~\ref{Tab:CubeTimes}. We find that within the variable length representation the traversal times are significantly reduced, as compared to the fixed length representation. Furthermore, also in the
current case of a second-order phase transition, the flat-var approach performs almost optimal, leading to a similar round-trip time than the 
opti-var method, alert with a mild difference between $\tau_{up}$ and $\tau_{down}$, which become equal in the opti-var approach.

In the following, we concentrate on the opti-var method, and compare results obtained using either  multi-cluster or single-cluster 
updates 
in the SSE loop updates. The corresponding round-trip times in Tab.~\ref{Tab:CubeTimes} already indicate that using single cluster updates, the random walker's dynamics suffers from a slowing-down in its dynamics. 

\begin{figure}
  \includegraphics[width=\columnwidth]{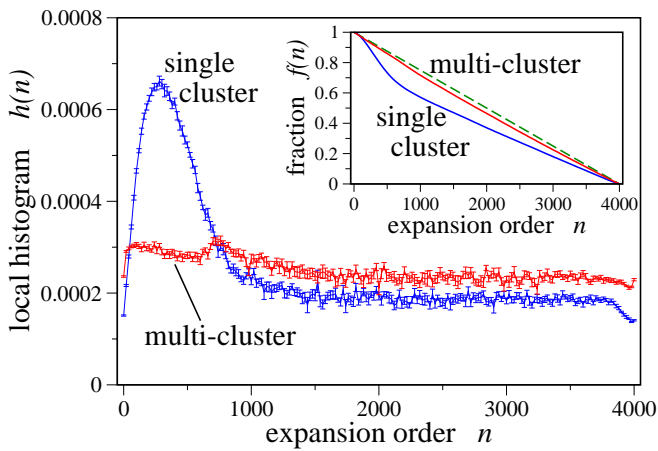}
  \caption{
  Sampled local histograms $h(n)$ for the optimized ensemble
  using single and multi-cluster updates in  a variable length representation stochastic series 
  expansion of the spin-1/2 Heisenberg model on a cubic lattice 
  with $N_s=8^3$ sites.
  The inset shows the respective fractions of random walkers diffusing
  from lowest to highest expansion order. For comparison, a linear decrease $f(n)=1-n/\Lambda$ is denoted by a dashed line.
  }
  \label{Fig:CubeHistogram}
\end{figure}

In Fig.~\ref{Fig:CubeHistogram}, we present  the histograms for both approaches, the inset showing the corresponding fractions.
We find that using multi-cluster updates, the opti-var method does not suffer any significant slowing-down due to the presence of the phase 
transition, as seen from the almost linear decreasing fraction $f(n)$. There is only a mild curvature visible;  similarly, the  histogram $h(n)$ 
does  not exhibit pronounced features for the multi-cluster approach. Using single-cluster updates is  less efficient in the critical 
temperature regime, and the opti-var method compensates for this inefficiency by shifting resources into the corresponding $n$-range. This 
illustrates, that the optimized ensemble method in the variable length representation, in combination with efficient cluster update schemes,   
allows for efficient simulations of a second-order phase transition: the cluster-updates eliminate the critical slowing-down, so that in the 
optimized ensemble, a uniform diffusivity can be achieved. Furthermore, we find that also in this case of a second-order phase transition, the 
flat histogram approach in the variable length representation performs almost optimal, showing very similar round-trip times than the optimal 
ensemble method.


\subsection{Thermal first-order phase transition}
In this section, we apply our analysis to a system undergoing a first-order thermal phase transition. For this purpose, we study  a system of 
hardcore bosons on a two-dimensional square lattice,
with next-nearest neighbor
repulsion~\cite{schmid}, described by the Hamiltonian
\begin{equation}
H=-t\sum_{\langle i,j \rangle}(a^{\dagger}_ia_j+a^{\dagger}_ja_i) +
V_2\sum_{\langle \langle i,k \rangle \rangle} n_in_k - \mu \sum_i n_i
\label{eq:Ham.hb}
\end{equation}
where $a^{\dagger}_i$ ($a_i$) creates (annihilates) a hardcore boson
at site $i$, $n_i$ is the density at this site, $t$ the hopping
amplitude between nearest neighbor sites, $V_2>0$ the next-nearest
neighbor repulsion, and $\mu$ the chemical potential. 
In the following, we consider the case of $t/V_2=0.45$ for a half-filled lattice at $\mu/V_2=2$. For this parameter values, the model 
exhibits a first-order phase transition at $T_c \approx 0.4 \, V_2$ from a high-temperature normal fluid phase to a low-temperature smectic phase with stripe order~\cite{schmid}.
In Ref.~\cite{QWL}, the improvement of the flat-fixed method over
conventional SSE simulations was quantified by studying the number of MC steps required to flip the directional orientation of the 
stripes in the smectic phase upon crossing  $T_c$. Here, we revisit this model in order to analyze the behavior of the opti-var method
for a system undergoing a first-order phase transition. We consider finite systems of $N_s=L\times L$ lattice sites with periodic boundary conditions, and scale the cutoff $\Lambda=20L^2$ in order to cover in each case a temperature range down to $T\approx 0.2 V_2<T_c$.
Since  for this system exact results for $g_{ex}(n)$ are not accessible, we performed a similar procedure as presented in Sec.~\ref{Sec:Cube}
in order to reach an optimized ensemble and calculate $g(n)$. For the non-local part of the QMC updates, we employed the directed loop method~\cite{DirectedLoops}, where the number of worms $N_W(n)$ constructed at a given expansion order $n$ was kept constant during the fixed weights simulations. The appropriate value of $N_W(n)$  for each expansion order $n$ was determined during the preceding Wang-Landau part of the simulations, such that on average $2n$ bond operators were visited during the non-local update part of each MC step~\cite{DirectedLoops}.

\begin{figure}
  \includegraphics[width=\columnwidth]{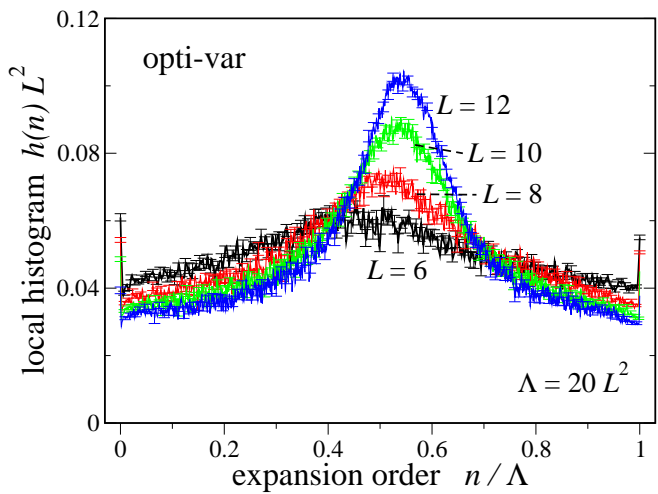}
  \caption{
  Optimized local histograms for a stochastic series expansion in the inverse
  temperature for a system of hardcore bosons with a repulsive next-nearest
  neighbor interaction on a square lattice of size $L \times L$.
  With increasing system size a peak evolves in the rescaled histogram as the system 
  undergoes a weak thermal first-order transition corresponding to a characteristic order
  $n_c \approx 0.55 \Lambda$ (with a maximal expansion order of
  $\Lambda = 20 L^2$).
  }
  \label{Fig:StripesHistogram}
\end{figure}
Figure~\ref{Fig:StripesHistogram} shows the local histograms $h(n)$ in the optimized ensemble method opti-var for different system sizes. 
In contrast to the flat-var method, 
we find that the local histograms now develop  peaks at an expansion order $n_c\approx 0.55 \Lambda$, indicative of the first-order phase transition in the corresponding temperature regime. The opti-var method shifts resources into this region of the $n$-range, where the local diffusivity, c.f. Eq.~(\ref{Eq:OptimalHistogram}), shows a suppression that becomes more pronounced upon increasing the system size. In contrast to the previous cases, we expect such a behavior from the first-order nature of the phase transition in the current case: The optimized ensemble  compensates for the inefficiency of the SSE updates to tunnel between the two coexisting phases at this first-order phase transition.

Let us thus compare the performance of the opti-var method to the flat-var approach.
Tab.~\ref{Tab:StripesTimes} shows the corresponding round-trip times for various system sizes. We find that the opti-var method indeed reduces the round-trip time of the random walkers as compared to the flat-var method. The improvement becomes more pronounced with increasing system size, as expected from the increasingly sharpening of the local histogram in Fig.~\ref{Fig:StripesHistogram}. Increasing the sampling inside the transition region, the opti-var method thus enhances  the overall diffusion of the random walker by 
pushing it towards this bottleneck.
\begin{table}
\begin{tabular}{|l|l|l|}
\hline
L & $\tau^{opti-var}_{\rm round}/10^6$&$\tau^{flat-var}_{\rm round}/10^6$\\
\hline
6 & 0.465(2) & 0.479(7)\\
8 & 1.848(7) & 2.05(4)  \\
10 & 5.41(2) & 5.60(3)  \\
12 &12.93(5) & 16.5(2)  \\
\hline
\end{tabular}
\label{Tab:StripesTimes}
\caption{Traversal times of random walkers diffusing from lowest order
  to highest orders in a stochastic
  series expansion in the inverse temperature of hardcore bosons with a repulsive next-nearest
  neighbor interaction on a square lattice of size $L \times L$.
  Data is shown for  variable length operator
  string in the optimized (opti-var) and flat histogram (flat-var) ensemble, in units of $10^6$ operator insertion/removal attempts. 
  The statistical error, estimated by performing 16 independent runs in each case is specified for the least significant digit.}
\end{table}

\subsection{Spinflop transition}

\begin{figure}
  \includegraphics[width=\columnwidth]{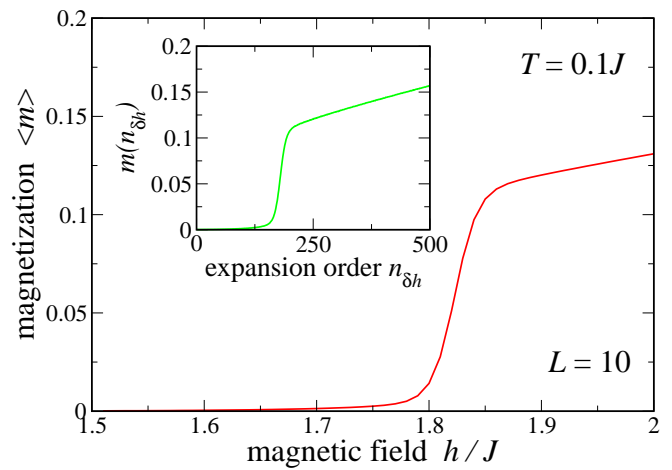}
  \caption{
  Magnetization vs. magnetic field 
  for the spin-1/2 XXZ model on a square lattice 
  with $10\times 10$ sites at $T/J=0.1$.  
  The inset shows the magnetization $ m(n_{\delta h})$ measured at expansion order $n_{\delta h}$ during
  an stochastic series expansion simulation using the optimized ensemble in the variable length operator string representation. 
  }
  \label{Fig:SpinFlopMagnetization}
\end{figure}

Thus far, we performed series expansions in the inverse temperature. Finally, we discuss results obtained from performing
a perturbation expansion in a model parameter~\cite{QWL}. In particular, we consider the spin-1/2 XXZ model
\begin{equation}
H=J\sum_{<i,j>} \left[ S^x_i S^x_j + S^y_i S^y_j + \Delta S^z_i S^z_j\right] - h \sum_i S^z_i,
\end{equation}
on a square lattice with $N_s=L\times L$ sites with periodic boundary conditions, 
in a finite magnetic field $h$ at an easy-axis anisotropy $\Delta=1.5$. The model shows a spinflop transition at $h_c\approx 1.83J$, where the magnetization $m$ jumps from $m=0$ for $h<h_c$ to a finite value of $m\approx 0.12$, c.f. Fig.~\ref{Fig:SpinFlopMagnetization}. In the following, we analyze the behavior of the opti-var method
as applied to this strongly first-order  phase transition. We fix the temperature  $T=0.1J$, and perform an expansion in the magnetic field. For this purpose, we decompose $h=h_0+\delta h$, where $h_0=1.5$, and introduce separate bond operators for the $\delta h$ contribution to the magnetic field.

In the SSE we perform an expansion
\begin{equation}
Z=\sum_{n_{\delta h}=1}^{\Lambda_{\delta h}} g( n_{\delta h} ) (\delta h)^{n_{\delta h}}
\end{equation}
in the parameter $\delta h$, from which  the expansion coefficients $g( n_{\delta h} )$,  $n_{\delta h}=1,...,\Lambda_{\delta h}$ are estimated~\cite{QWL}.  In addition, we also measure after each full MC update step the total magnetization $m=\sum_{i=1}^{N_s} S^z_i$ of the system, and bin these values according to the current expansion order $n_{\delta h}$. This way, we obtain a magnetization histogram $m(n_{\delta h})$. The inset of Fig.~\ref{Fig:SpinFlopMagnetization} shows an example of a magnetization histogram $m(n_{\delta h})$ for $n_{\delta_h}=0,...,\Lambda_{\delta h}=500$. The magnetization histogram is used to calculate the field dependent magnetization from
\begin{equation}
\langle m\rangle=\frac{1}{Z}\sum_{n_{\delta h}=1}^{\Lambda_{\delta h}} m(n_{\delta h} ) g( n_{\delta h} ) (\delta h)^{n_{\delta h}},
\end{equation}
such as shown in the main part of Fig.~\ref{Fig:SpinFlopMagnetization}. 

Similar to the previous cases, we used a two-step procedure to obtain an optimized ensemble.
We first performed 8 Wang-Landau sampling steps starting from a uniform assignment $g(n_{\delta h})=1, n_{\delta_h}=0,...,\Lambda_{\delta h}=500$, in order to obtain a broad-histogram ensemble. Then, we performed 4 feedback iteration batches, where the number of MC steps was increased by a factor of 2 after each feedback step. The magnetization histogram is recorded only during the final step. Again, we performed 16 independent simulations for statistical averaging over the final results.

\begin{figure}
  \includegraphics[width=\columnwidth]{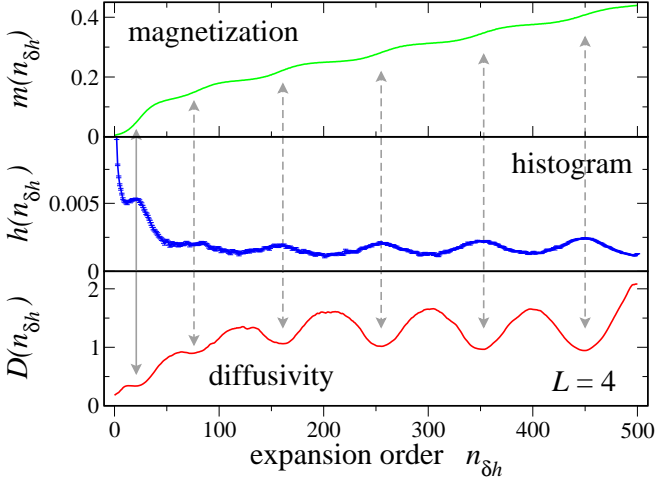}
  \caption{
  Magnetization, optimized local histogram and diffusivity of random walkers in a stochastic series 
  expansion for the spin-1/2 XXZ model on a square lattice with $4\times 4$ sites.
  The diffusivity
  shows suppressions for expansion orders where the magnetization shows 
  a larger increase. 
  The optimized ensemble compensates for the respective slowing-down by shifting 
  additional resources towards these expansion orders (see middle panel).      
  }
  \label{Fig:SpinFlopDiffusivity}
\end{figure}
Let us first consider the results for a small system with $L=4$ in the opti-var method. In the top panel of Fig.~\ref{Fig:SpinFlopDiffusivity}, 
the 
magnetization histogram for this system is shown. It displays more structure than the  magnetization histogram in Fig.~\ref{Fig:SpinFlopMagnetization} for the $L=10$  system. In particular, apart from a first increase of $m(n_{\delta h} )$ near $n_{\delta h}\approx 20$ (indicated by the solid arrowed vertical line), a series of additional 
increases in $m(n_{\delta h} )$ are found at higher expansion orders (dashed arrowed vertical lines). While the increase near $n_{\delta h}\approx 20$, followed by a mild plateau of $m\approx 0.125$ corresponds to the  spinflop transition, the  features at larger $n_{\delta h}$ relate to the discrete magnetization steps on a finite lattice, which for $L=4$ amount to $\Delta m=1/L^2=0.0625$. Indeed, for $T=0$ the magnetization process is a series of steps, with jumps of the magnetization by $\Delta m$. At $T=0.1J$, these steps are less pronounced, but still  visible in Fig.~\ref{Fig:SpinFlopDiffusivity}. In the plot of the diffusivity $D(n_{\delta h})$ for the optimized ensemble, the reduced efficiency of the SSE algorithm close to  magnetization jumps~\cite{DirectedLoops} is clearly visible (see bottom panel of Fig.~\ref{Fig:SpinFlopDiffusivity}). As seen from the optimized histogram, shown in the middle panel of Fig.~\ref{Fig:SpinFlopDiffusivity}, the optimized ensemble shifts resources into these regions. For the magnetization increase near  $n_{\delta h}\approx 20$, related to the  spinflop transition, these features in $D(n_{\delta h})$ and $H(n_{\delta h})$  are also visible, but less dominant.
However, upon increasing the system size, this slowing-down of the simulation near the spinflop transition becomes more pronounced, and dominates
the dynamics of the random walker.

\begin{figure}
  \includegraphics[width=\columnwidth]{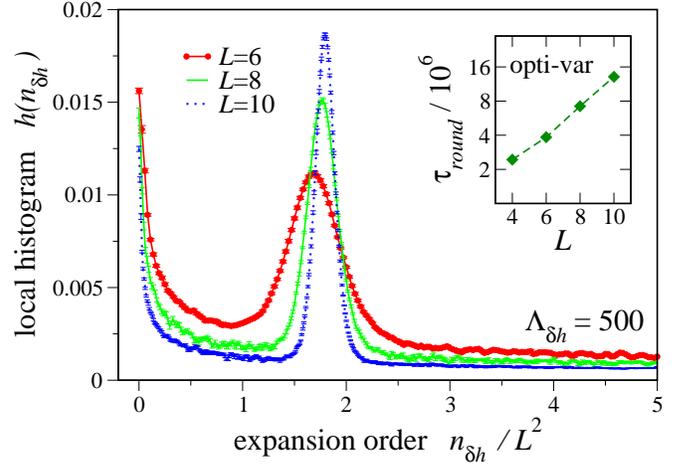}
  \caption{
  Sampled local histograms for the optimized ensemble in a stochastic series expansion
  for the spin-1/2 XXZ model on a square lattice with $L\times L$ sites. 
  The perturbative expansion at temperature $T=0.1 J$ is carried out in 
  terms of the magnetic field $\delta h$.
  A pronounced peak in the histogram evolves at a characteristic order of the
  spinflop transition which scales linearly with the linear 
  system size $L$. The inset shows the system size dependence of the round-trip time in a linear-log plot.
  }
  \label{Fig:SpinFlopHistogram}
\end{figure}

Figure~\ref{Fig:SpinFlopHistogram} shows the histogram for the optimized ensemble for larger systems. A pronounced peak develops 
around an expansion order, characteristic of the spinflop transition for each system size. This feature dominates the optimized histogram, and the additional magnetization steps at larger expansion orders become less relevant, as the magnetization increases smoothly for $h>h_c$. Both the position and the height of the spinflop peak scale linearly with the linear system size $L$. The width of the peak does not change significantly upon increasing $L$, indicating that the spinflop transition becomes increasingly sharp in $h$, as $L$ increases.

\begin{figure}
  \includegraphics[width=\columnwidth]{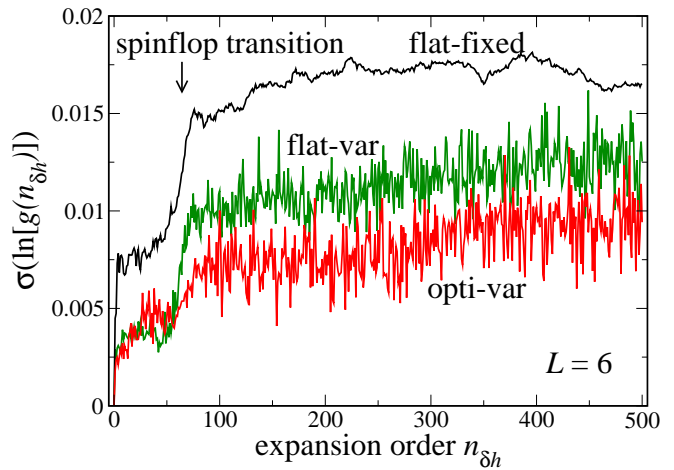}
  \caption{
  Averaged statistical errors of the expansion coefficients $g(n_{\delta h})$ calculated in a 
  stochastic series expansion for the spin-1/2 XXZ model on a square lattice 
  with $6\times 6$ sites. 
  The statistical errors are averaged over 16 independent simulations for
  a fixed CPU time.
  Results are shown for the flat histogram ensemble with fixed length operator
  string (flat-fixed), variable length operator string (flat-var) and the optimized ensemble
  with variable length operator string (opti-var).
  The characteristic expansion order, corresponding to  
  the spinflop transition is indicated by the arrow. 
  }
  \label{Fig:SpinFlopError}
\end{figure}

In Tab.~\ref{Tab:SinFlopTimes}, we compare the traversal times of the opti-var algorithm to the flat histogram methods flat-fixed and flat-var 
for the $L=6$ system. In the fixed length flat-fixed method,  strong differences in $\tau_{\rm down}\approx 7 \tau_{\rm up}$ are found, while in 
both variable length approaches (flat-var and opti-var) the random walker moves  balanced ($\tau_{\rm down}\approx \tau_{\rm up}$). Moreover, in 
both variable length methods, the round-trip time is significantly reduced as compared to the flat-fixed method, with the optinal ensemble  
resulting in the largest overall speedup. 
\begin{table}
\begin{tabular}{|l|l|l|l|}
\hline
Method & $\tau_{\rm up}/10^6$ & $\tau_{\rm down}/10^6$ & $\tau_{\rm round}/10^6$ \\
\hline
flat-fixed & 3.81(3)  & 21.7(3)  & 25.5(3) \\
flat-var & 2.31(5)  & 2.30(3) & 4.62(3) \\
opti-var & 1.95(2)  & 1.92(2) & 3.87(2)  \\
\hline
\end{tabular}
\label{Tab:SinFlopTimes}
\caption{Traversal times of random walkers diffusing from lowest order
  to highest orders in a stochastic
  series expansion in the magnetic field close to the spinflop transition on a $6\times 6$ lattice.
  Data is shown for the flat histogram ensemble with fixed length operator
  string (flat-fixed) and variable length operator string (flat-var), and the optimized ensemble for the variable length operator string representation (opti-var),
  in units of $10^6$ operator insertion/removal attempts. The statistical error, estimated by performing 16 independent runs in each case is specified for the least significant digit.}
\end{table}
This improved simulation dynamics leads to an equivalent reduction of the statistical errors in the expansion coefficients $g(n_{\delta h})$. We  performed in each case 16 fixed CPU time simulations in order to assess the standard statistical error in the $g(n_{\delta h})$. The results for the $L=6$ system are shown in Fig.~\ref{Fig:SpinFlopError}. While the flat-fixed method suffers from a strong increase of the statistical error over the $n$-range that relates to the spinflop transition, the optimized ensemble method performs more uniformly over the whole $n$-range. We  find a  uniform reduction of the statistical error in higher $n$-range; the size of this reduction fits well to the square-root of the round-trip time fraction shown in Tab.~\ref{Tab:SinFlopTimes}. This relation between the statistical error and the inverse square-root of the round-tip time was  observed also in classical systems~\cite{OptimizedEnsembles}. 

Finally, we assess, how the optimal ensemble method scales upon increasing the system size.
In the inset of Fig.~\ref{Fig:SpinFlopHistogram}, the round-trip time $\tau_{\rm round}$ is shown vs. the linear system size $L$ for the opti-var method. The observed dependence fits well to an exponential increase of $\tau_{\rm round}$ with $L$, indicative of the strong first-order nature of the spinflop transition. 
While the optimized ensemble does thus still suffer from an exponential performance reduction, it provides 
a performance improvement over the flat-histogram method, also in this case.

%% file: appendix.tex
In the following, we prove detailed balance for the variable length operator string representation
update scheme of Sec.~\ref{Sec:VariableLength},
thereby extending previous work by D.~C. Handscomb~\cite{Handscomb} 
and A.~W. Sandvik~\cite{SandvikHubbard}.

In order to obtain the correct equilibrium distribution of configurations, it is sufficient to prove that the transition probabilities from one state to the next satisfy detailed balance:
\begin{equation}\label{dbsplit}
w(C)\cdot p(C\to C') = w(C')\cdot p(C' \to C), 
\end{equation}
with $w(C)$ the weight of a configuration $C$ and $p(C\to C')$ the transition probability from $C$ to $C'$. Here we prove detailed balance for the variable length operator string
representation \emph{per sweep}. The main idea of the proof is that for each sequence of operator insertions and removals in a sweep we can construct an inverse sequence, and these satisfy detailed balance. 
A configuration $C$ with $n$ operators is defined as a sequence of length $2n+1$ consisting of $n$ operators and $n+1$ gaps, including one at the beginning and one at the end
of the string. Configurations during a sweep contain an additional ``pointer''$k, 1\leq k\leq 2n+1,$ that labels the position at which we currently perform insertion and removal operations. While the number
of operators of the intermediate configurations between a sweep may grow and shrink, the pointer k increases by zero (removal) or two (insertion) with each step. We label
such intermediate configurations as $C_n(k).$

\begin{figure}[t!]
\begin{center}
\includegraphics[width=\columnwidth]{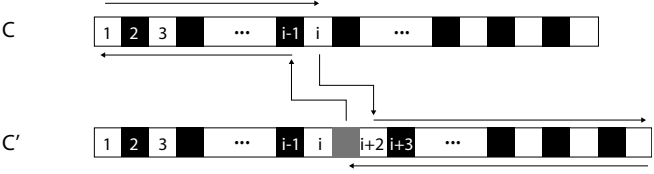}
\caption{Illustration of the transition $C\to C'$ by inserting an operator at position $i$. In our choice of indexing the operators, after a successful insertion all remaining operators shift indices by $2$, i.e.. $H_{b_k}$ in $C$ is identical to $H_{b_{k+2}}$ in $C'$ for all $k>i+1$. The arrows illustrated the positions of the pointers $k$  for intermediate configurations. All rejected update probabilities cancel except for the insertion/removal probabilities of the new operator (marked grey).}\label{figsticksdb1}
\end{center}
\end{figure}

For the sake of simplicity we first look at the insertion of only one operator at position $k$ during a sweep, as illustrated in Fig. \ref{figsticksdb1}. Our configuration $C$ containing $n$ operators and $n+1$ gaps is thus changed into a 
configuration $C'$ containing $n+1$ operators and $n+2$ gaps, consecutively labeled by an index $i, 1 \leq i \leq 2n+1$.
We label the probability of inserting any operator into a gap of an operator string of length n as $p_{\text{ins}}^n.$ Note that this probability is independent of the position of pointer k 
and depends only on n. With $p_{\text{ins}}^n(H_b)$ the probability of inserting operator $H_b$ and $p_{\text{rem}}^n(k)$ the one for removing the operator at position k (even) we obtain 

\begin{align}
p(C \to C') &= \left( 1- p_{\text{ins}}^n\right)^{\frac{i-1}{2}} \cdot \prod_{j=1}^{(i-1)/2} \left(1-p_{\text{rem}}^n(2j)\right) \\ \nonumber &\cdot \left[ \frac{1}{N_d} p_{\text{ins}}^n(H_{b_1})\right] \\ \nonumber &\cdot  \left(1-p_{\text{ins}}^{n+1}\right)^{\frac{2(n+1)+1-i}{2}} 
  \cdot \prod_{j=\frac{i+3}{2}}^{\frac{2(n+1)}{2}} (1-p_{\text{rem}}^{n+1}(2j)).
\end{align}
$N_d$ denotes the total number of bond operators. For computing the probability of an operator removal we take configuration $C'$ and traverse it in reverse order. This corresponds to reordering the operator string back to front, but since our algorithm
constructs only diagonal (thus commuting) operators, this reordering is legal. The transition probability 
\begin{align}
p(C' \to C) &= \prod_{j=\frac{i+3}{2}}^{\frac{2(n+1)}{2}} (1-p_{\text{rem}}^{n+1}(2j)) \cdot \left(1-p_{\text{ins}}^{n+1}\right)^{\frac{2(n+1)+1-i}{2}} \\ &\cdot p_{\text{rem}}^{n+1}(H_{b_1}) \nonumber \\ \nonumber
&\cdot \left(1- p_{\text{ins}}^n\right)^{\frac{i-1}{2}}   \cdot \prod_{j=1}^{(i-1)/2} \left(1-p_{\text{rem}}^n(2j)\right)
\end{align}
shows that for only one operator insertion per sweep all the rejected insertion and removal attempts cancel and detailed balance is fulfilled for our choice of operator insertion
and removal probabilities, see equations (\ref{insprob}), (\ref{remprob}):
\begin{equation}
\frac{p(C\to C')}{p(C' \to C)}
 = \frac{\frac{1}{N_d} \min\left[1, \frac{N_d \beta \langle \alpha(p) | H_{b_1} | \alpha(p-1)\rangle}{n+1}\right]}{\min\left[1, \frac{n+1}{N_d \beta \langle \alpha(p) | H_{b_1} | \alpha(p-1)\rangle}\right]}
= \frac{w(C')}{w(C)}.
\end{equation}
The removal of one operator works analogously.

\begin{figure}[t!]
  \begin{center}
   \includegraphics[width=\columnwidth]{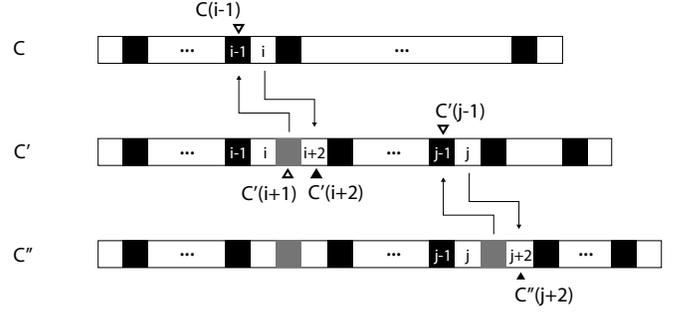}
   \caption{The transition from $C$ to $C''$ via an intermediate visit to $C'$. The two newly inserted    
   operators are marked grey and the arrows denote some of the intermediate configurations 
   encountered in the process in forward (filled arrows) and backward (empty arrows) direction. }
   \label{figsticksdb2}
  \end{center}
\end{figure}

In a next step we consider a sweep with two insertion updates, see Fig. \ref{figsticksdb2}. Again, we start with a configuration $C$ with $n$ operators. We add an operator at position $i$ as before, resulting
in a configuration $C'_{n+1}(i+2)$ of $n+1$ operators and a pointer at gap $i+2.$ We then proceed to a position $j, i+2\leq j\leq 2n+1,$ where we try another insertion. Once accepted we arrive
at a configuration $C''_{n+2}(j+2).$ 

Since the transition probabilities between intermediate configurations  depend on the direction of traversal, we denote them by arrows $\overrightarrow{p} $ for the forward or $\overleftarrow{p}$ for the backward direction. We need to prove $p(C\rightarrow C'')/p(C''\rightarrow C) = w(C'')/w(C).$
As we have to visit the intermediate states $C'_{n+1}(i+2)$ and $C''_{n+2}(j+2)$ in order to arrive at the final state $C'',$ we can split up the transition probabilities:

\begin{align}\label{eqccss}
 p(C \to C'') =& \overrightarrow{p}(C \to C'_{n+1}(i+2))\\ \nonumber \cdot& \overrightarrow{p}(C'_{n+1}(i+2)\to C''),\\
 p(C'' \to C) =& \overleftarrow{p}(C'' \to C'_{n+1}(i+1))\\ \nonumber \cdot& \overleftarrow{p}(C'_{n+1}(i+1) \to C).
\end{align}
As before, probabilities cancel when traversing operators and gaps in reverse order: 
\begin{equation}\label{eqgp1}
\overrightarrow{p}(C''_{n+2}(j+2) \to C'') = \overleftarrow{p}(C'' \to C''_{n+2}(j+1)).
\end{equation}
In order to establish detailed balance, let us perform one insertion, then visit state $C'$ with one more operator as before at the right boundary and bounce back to where we were (this move cancels), and proceed with the insertion. This will
be balanced by a removal move, a move to the right and back, and another removal move:
\begin{align}
\frac{p(C\to C'')}{p(C''\to C)} =& \frac{p(C\to C')}{p(C' \to C)} \frac{\overleftarrow{p}(C' \to C'_{n+1}(i+1))}{\overrightarrow{p}(C'_{n+1}(i+2)\to C')}\\ \nonumber \cdot &\frac{ \overrightarrow{p}(C'_{n+1}(i+2)\to C'')}{ \overleftarrow{p}(C''\to C'_{n+1}(i+1))} \cr
=& \frac{w(C')}{w(C)} \cdot \frac{ \overrightarrow{p}(C'_{n+1}(i+2)\to C'')}{ \overleftarrow{p}(C''\to C'_{n+1}(i+1)) }\cr
=& \frac{w(C')}{w(C)} \cdot \frac{p(C' \to C'')}{p(C''\to C')}\cdot \frac{\overleftarrow{p}(C'_{n+1}(i+1)\to C')}{\overrightarrow{p}(C' \to C'_{n+1}(i+2))}\cr
=& \frac{w(C')}{w(C)} \cdot \frac{w(C'')}{w(C')}
= \frac{w(C'')}{w(C)} \,.
\end{align}
Again, the removal operation is exactly the same.

We now construct a setup that allows us to treat more than one update per sweep. For this we need to take into account that there may be more than one sequence of intermediate insertion and 
removal operations
$S_j=\{C_{j1}, \cdots, C_{jp}\}$ that change an initial configuration $C$ via intermediate configurations in $S_j$ into a final configuration $C'$. We proceed by constructing a reverse sequence $S'$ for
each $S$ and show that most probabilities cancel. The detailed balance condition
$p(C \stackrel{S_j}{\rightarrow} C')/p(C' \stackrel{S'_j}{\rightarrow}C) = \frac{w(C')}{w(C)}$ 
for each one of these sequences is established in the same manner as above, with a ``visit to the right boundary'' after 
each successful insertion or removal move.

Again, we start with a configuration $C$ with $n$ operators. We add or remove the first operator to arrive at the first intermediate state of $S_j$, $C_{1j}$ at position $i_{1j}$ as before.
We then transverse to the right of that operator string by increasing i, inserting and removing operators according to $S_j$ and finally arriving
at a configuration $C''$ in one sweep. The reverse sequence $S'_j$ is defined by the intermediate states of $S_j$ in reverse order. Analogously to (\ref{eqccss}) we obtain
\begin{align}
p(C \stackrel{S_j}{\rightarrow}C')/p(C' \stackrel{S'_j}{\rightarrow} C) = \frac{w(C')}{w(C)},
\end{align}
and therefore
\begin{align}
&\frac{p(C \rightarrow C')}{p(C' \rightarrow C)} =  \frac{\sum_j p(C\stackrel{S_j}{\rightarrow} C')}{\sum_j p(C'\stackrel{S'_j}{\rightarrow} C)} \\ \nonumber = &\frac{\sum_j w(C')/w(C) p(C' \stackrel{S_j}{\rightarrow} C)}{\sum_j p(C'\stackrel{S'_j}{\rightarrow} C)} = \frac{w(C')}{w(C)}.
\end{align}
This framework of traversing the operator string to the end and back for each operator is only required for the proof. During the calculation we just keep a variable length operator string and
a pointer k that marks the position at which the next insertion or removal operation is performed according to the probabilities in equations $(\ref{insprob}), (\ref{remprob}).$